\documentclass[11pt,twoside]{article}
\usepackage{./asp2014}

\aspSuppressVolSlug
\resetcounters

\begin{document}

\title{B[e] Supergiants' circumstellar environment: disks or rings?}
\author{G. Maravelias$^1$, M. Kraus$^{1,2}$, A. Aret$^2$, L. Cidale$^{3,4}$, M. L. Arias$^{3,4}$, M. Borges Fernandes$^5$
\affil{$^1$Astronomick\'y \'ustav AV\v{C}R, v.v.i., Ond\v{r}ejov, Czech Republic; \email{maravelias@asu.cas.cz} \\
$^2$Tartu Observatory, T{\~o}ravere, Estonia \\
$^3$Facultad de Ciencias Astron\'omicas y Geof\'isicas, UNLP, La Plata, Argentina \\
$^4$Instituto de Astrof\'isica de La Plata, La Plata, Argentina \\
$^5$Observat\'orio Nacional, Rio de Janeiro, Brazil }
}

\paperauthor{G.~Maravelias}{maravelias@asu.cas.cz}{}{Astronomick\'y \'ustav AV\v{C}R, v.v.i.}{Stellar Dept.}{Ond\v{r}ejov}{}{25165}{Czech Republic}
\paperauthor{M.~Kraus}{kraus@asu.cas.cz}{}{Astronomick\'y \'ustav AV\v{C}R, v.v.i.}{Stellar Dept.}{Ond\v{r}ejov}{}{25165}{Czech Republic}
\paperauthor{M.~Kraus}{kraus@to.ee}{}{Tartu Observatory}{}{T\~oravere}{Tartumaa}{61602}{Estonia}
\paperauthor{A.~Aret}{aret@to.ee}{}{Tartu Observatory}{}{T\~oravere}{Tartumaa}{61602}{Estonia}
\paperauthor{L.~Cidale}{lydia@fcaglp.fcaglp.unlp.edu.ar}{}{Facultad de Ciencias
Astron\'omicas y Geof\'isicas, Universidad Nacional de La Plata}{Departamento de Espectroscop\'ia Estelar}{La Plata}{}{}{Argentina}
\paperauthor{L.~Cidale}{lydia@fcaglp.fcaglp.unlp.edu.ar}{}{Instituto de Astrof\'isica de La Plata, CCT La Plata, CONICET-UNLP}{}{La Plata}{}{}{Argentina}
\paperauthor{M.L.~Arias}{mlaura@fcaglp.fcaglp.unlp.edu.ar}{}{Facultad de Ciencias
Astron\'omicas y Geof\'isicas, Universidad Nacional de La Plata}{Departamento de Espectroscop\'ia Estelar}{La Plata}{}{}{Argentina}
\paperauthor{M.L.~Arias}{mlaura@fcaglp.fcaglp.unlp.edu.ar}{}{Instituto de Astrof\'isica de La Plata, CCT La Plata, CONICET-UNLP}{}{La Plata}{}{}{Argentina}
\paperauthor{M.~Borges~Fernandes}{borges@on.br}{}{Observat\'orio Nacional}{}{Rio de Janeiro}{}{}{Brazil}

\begin{abstract}
B[e] Supergiants are a phase in the evolution of some massive stars for which we have observational evidence but no predictions by any stellar evolution model. The mass-loss during this phase creates a complex circumstellar environment with atomic, molecular, and dust regions usually found in rings or disk-like structures. However, the detailed structure and the formation of the circumstellar environment are not well-understood, requiring further investigation. To address that we initiated an observing campaign to obtain a homogeneous set of high-resolution spectra in both the optical and NIR (using MPG-ESO/FEROS, GEMINI/Phoenix and VLT/CRIRES, respectively). We monitor a number of Galactic B[e] Supergiants, for which we examined the [OI] and [CaII] emission lines and the bandheads of the CO and SiO molecules to probe the structure and the kinematics of their formation regions. We find that the emission from each tracer forms either in a single or in multiple equatorial rings.
\end{abstract}

\section{Introduction}

One of the main groups of objects that exhibit the B[e] phenomenon are B[e] Supergiants (B[e]SGs; \citealt{Lamers1998}), which experience strong aspherical mass loss. This results in a complex circumstellar environment that combines atomic, molecular, and dust regions. The traditional picture assumes a continuous outflowing disk, starting very close to the star, where the forming regions of permitted and forbidden lines of neutral and ionized elements reside and the dust regions form further outside \citep{Zickgraf1985}. However, this picture has changed significantly by observations over the last decade, as detached Keplerian rotating disks or rings of circumstellar material have been identified \citep[e.g.][]{Millour2011,Aret2012,Cidale2012,Wheelwright2012a,Oksala2013,Kraus2016}.

Nevertheless, these works focus on individual stars often observed with different instruments. Driven by the lack of a systematic survey performed with a homogeneous approach, we initiated a campaign to study the circumstellar environments of a large number of Galactic B[e]SGs, using high-resolution optical and NIR spectroscopy. The current contribution summarizes our findings regarding 6 selected Galactic B[e]SGs.

\section{Observations and data reduction}

We obtained optical spectra using the FEROS spectrograph, attached to the 1.52 m telescope for observations obtained in 1999, and to the 2.2 m telescope later on, at ESO (La Silla, Chile). FEROS is a high-resolution echelle spectrograph with fibers, covering a wide range 3600-9200 \AA\, at a spectral resolution of $R=55000$ (around 6000 \AA). The observations were reduced with the FEROS automatic pipeline. Telluric correction was applied with IRAF using a standard star observed on the same night, and whenever this was not possible we used telluric templates from other nights.
For the NIR spectra we used the Phoenix spectrograph, attached to the 8 m Gemini South telescope (Cerro Pachon, Chile), and the CRIRES instrument, attached to the 8.2 m UT1/VLT ESO telescope (Cerro Paranal, Chile). Phoenix is a cryogenic, long-slit spectrograph covering the 1-5 $\mu$m range at a spectral resolution of $R=50000-80000$. CRIRES is a similar instrument covering the 0.95-5 $\mu$m range at a spectral resolution of $R=50000-100000$. During observations a standard nodding strategy was followed to remove sky emission, and spectra from standard telluric stars were obtained each night. The data reduction and telluric correction were performed using IRAF software.

\section{Model fitting}

With high-resolution spectra we are able to resolve and study the kinematics of selected emission features, which probe different regions in the disk. In the NIR spectra we can detect the CO bands in emission, which originate from the hot inner edge of the molecular disk \citep[e.g.][]{Kraus2009,Liermann2010,Cidale2012,Oksala2013,Muratore2015}. At NIR, emission from SiO molecules can be also detected, in regions close to CO but at slightly lower temperatures ($T_{SiO}<T_{CO}<\sim5000\, K$; \citealt{Kraus2015}). In the optical spectra the set of optically thin emission lines displaying broadening are the [OI] $\lambda$5577 line, and the doublets [OI] $\lambda\lambda6300,6363$ and [CaII] $\lambda\lambda7291,7323$ (thereafter, the atomic gas). Earlier investigations have shown that in a Keplerian disk the [CaII] and the [OI] $\lambda$5577 lines form close to the star, and the [OI] doublet forms further out \citep{Kraus2010,Aret2012}.

Assuming a purely kinematical approach, the broadening of the aforementioned features can be modeled by a narrow rotating ring of gas. This velocity (de-projected, whenever the inclination angle is known) is convolved with a Gaussian component which is a combination of the instrument's spectral resolution ($\sim5.5-6.5\,\text{km}\,\text{s}^{-1}$ for FEROS), a typical thermal velocity ($\sim1-2\,\text{km}\,\text{s}^{-1}$ for the optical lines), and some random internal motion of the gas (a few $\text{km}\,\text{s}^{-1}$). In many cases this random motion may be larger, which we interpret as typical ring-width. 

Our approach in determining the rotational velocities is based on the following strategy: (i) as CO emission features are a direct evidence of gas presence, we first determine the rotational velocity for the CO ring, (ii) assuming that the optical emission lines originate from the same region as that of CO (in a homogeneous disk) we use the same velocity derived from CO, and we check if it results in sufficient broadening, (iii) if the observed profile is not matched then we need either to change this value or add more rings in our model. A more detailed example on this procedure is presented in Fig. \ref{fig1}, regarding emission features of CPD-52 9243. 

\articlefigure[width=.7\textwidth]{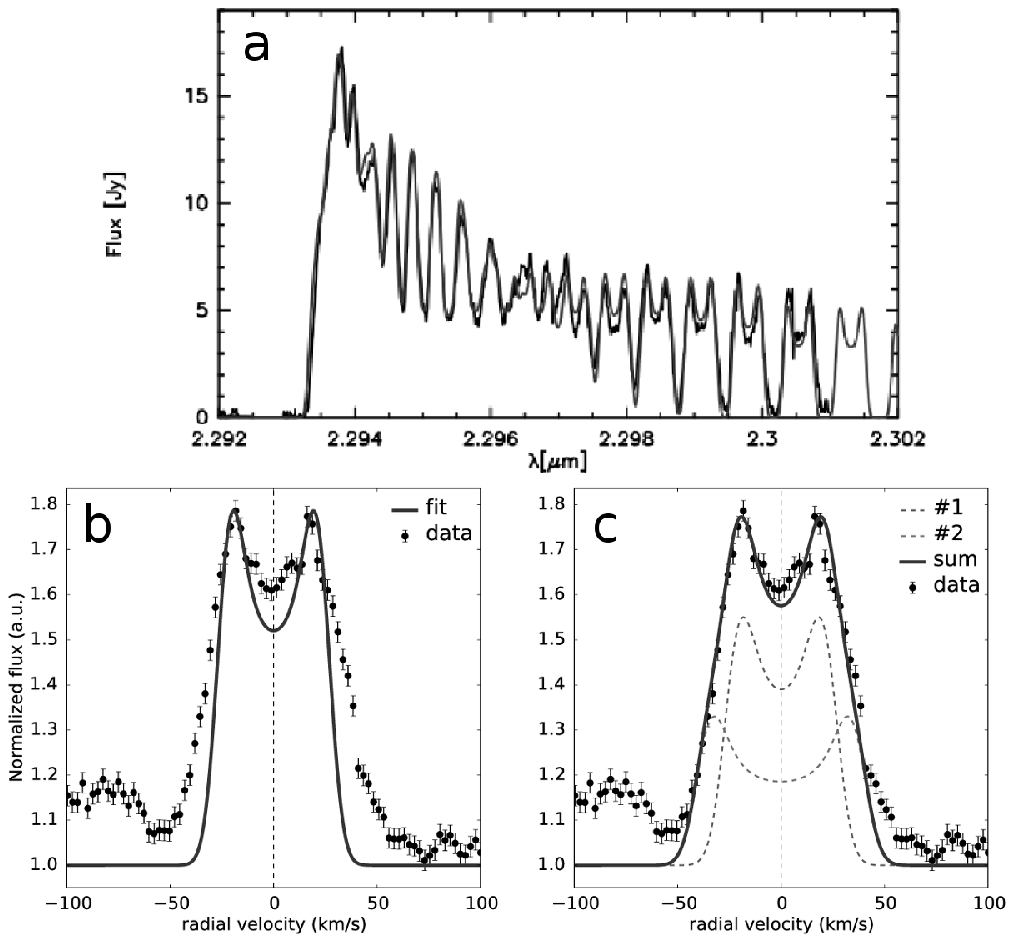}{fig1}{
An example of the model fitting process for CPD-52 9243: (a) Starting with the first CO bandhead emission feature (data and model shown as black and gray line, respectively) we determine the rotational velocity of CO ring at 33.5 $\text{km}\,\text{s}^{-1}$ (after \citealt{Cidale2012}), (b) Assuming the same rotational velocity with CO, i.e. the same emitting region, for the [CaII] $\lambda$7292 line, we see that the broadening is not enough to fit the observed profile (data shown as black dots, and the model as solid black line, with a rotational velocity of  $Vrot=33.5\pm2\,\text{km}\,\text{s}^{-1}$ and a Gaussian component of $Vg=8\pm1\,\text{km}\,\text{s}^{-1}$  ). (c) If we use two rings (dashed lines at $Vrot_1=36\pm1\,\text{km}\,\text{s}^{-1}$ and $Vg_1=9\pm1\,\text{km}\,\text{s}^{-1}$, $Vrot_2=52\pm1\,\text{km}\,\text{s}^{-1}$ and $Vg_2=9\pm1\,\text{km}\,\text{s}^{-1}$) then their total profile matches the observed one. Thus, we conclude that [CaII] forms in two separate regions with a typical ring-width of $\sim5\,\text{km}\,\text{s}^{-1}$ each.
}

\section{Results}

We summarize our results in Fig. \ref{fig2}, where we also "visualize" the structures according to the physical parameters for each object. We note that the [OI] $\lambda$5577 line was not detected in any object, and we refer to the [OI] $\lambda$6300 line only. 

For HD 327083 (RA=17:15:15.4, Dec=-40:20:06.7; J2000) we identify three rings. In the ring closer to the star (i.e. with the higher velocities at $\sim86-84\,\text{km}\,\text{s}^{-1}$) we find both molecular CO and atomic [OI] and [CaII] gas. Further outside we find a ring of SiO (at $78\,\text{km}\,\text{s}^{-1}$) and another one of atomic gas ([OI] and [CaII], at $\sim68-64\,\text{km}\,\text{s}^{-1}$). Due to the typical ring-width the CO and SiO rings could form a common region. HD 327083 is a binary system \citep{Wheelwright2012} that is located inside the CO ring, which means that the whole structure is circumbinary. 

Using IR spectroscopy and interferometry \cite{Cidale2012} found that CO emission in CPD-52 9243 (RA=16:07:01.9, Dec=-53:03:45.7; J2000) originates from a detached disk/ring, rotating at 36 $\text{km}\,\text{s}^{-1}$. By fitting our optical data we find two distinct rings at 53 and 36 $\text{km}\,\text{s}^{-1}$, respectively. The innermost ring consists of atomic gas only ([OI],[CaII]), while the outer ring is a mix of all tracers ([OI],[CaII],CO,SiO).

Similarly, IR interferometric observations on HD 62623 (3 Pup, RA=07:43:48.5, Dec=-28:57:17.3; J2000) revealed the presence of a detached hot ionized disk close to the star and a dusty disk further outside \citep{Millour2011}. Our results show the presence of two major emitting regions at 76 and $\sim53-50\,\text{km}\,\text{s}^{-1}$ (since the value of $48\,\text{km}\,\text{s}^{-1}$ for SiO is very close due to the typical ring-width, so not clearly separated). The inner ring consists only of atomic gas ([OI],[CaII]), while the outer one consists of both atomic and molecular gas ([OI],[CaII],CO,SiO). Given the new mass estimate by \cite{Aret2016} this picture is consistent with the interferometric results.

For Hen3-298 (RA=09:36:44.4, Dec=-53:28:00.0; J2000) we find two emitting regions at 22 and 19 $\text{km}\,\text{s}^{-1}$ (line-of-sights velocities due to unknown inclination angle). [CaII] is found closer to the star than [OI] and CO which coexist only slightly further out. However, due to the  large ring-width these two regions may not be different.

MWC 137 (RA=06:18:45.5, Dec=+15:16:52.3; J2000) has a very complex circumstellar environment. It is surrounded by an optical nebula (Sh 2-266) with many IR structures (see Kraus et al., this volume), and a jet \citep{Mehner2016}. A detached CO emitting ring has been detected at 84 $\text{km}\,\text{s}^{-1}$. From our optical spectra we find no atomic gas in the CO ring or closer to the star. On the contrary, we find a sequence of four distinct rings of [OI] at 70, 47, 31, 20 $\text{km}\,\text{s}^{-1}$ (line-of-sight velocities due to uncertain inclination angle). Moreover, no emission from [CaII] is present.

CPD-57 2874 (RA=10:15:21.9, Dec=-57:51:42.7; J2000) needs a larger number of rings to better fit its optical emission lines. We use 5 different rings with rotational velocities spanning the range 130-26 $\text{km}\,\text{s}^{-1}$. No gas is present closer to the star than the CO ring, which has been found to be detached \citep{DomicianodeSouza2011}. Our results show a sequence of multiple detached rings, which could indicate also a disk structure. We identify only [CaII] coexisting with CO, while SiO seems to form in a separate ring. Further outside these two molecular regions we find rings of atomic gas combining both [CaII] and [OI], except for the very last ring where only [OI] is found.

\articlefigure{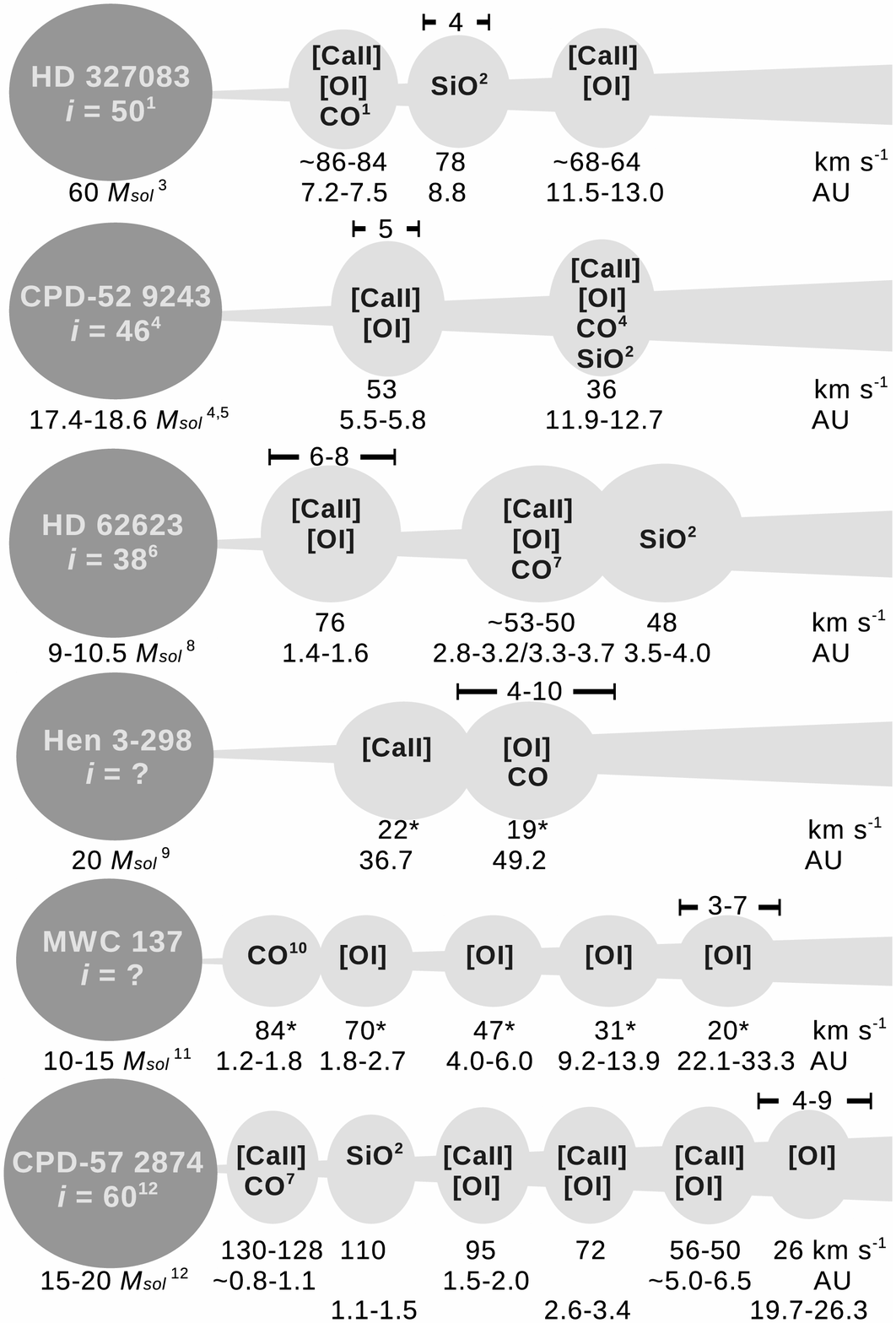}{fig2}{
Summary of identified rotational velocities by fitting the optical [OI] $\lambda$6300 and [CaII] $\lambda7293$ emission lines with narrow rings of material, which are shown as simple sketches (not to scale) according to the physical parameters of each system. A typical ring-width (in $\text{km}\,\text{s}^{-1}$) for each system is given as a distance indicator over a ring. Rotational velocities with a '*' sign indicate line-of-sight velocities due to unknown/uncertain inclination angles. [\textit{References}: $^1$Marchiano et al. (in prep.), $^2$\cite{Kraus2015}, $^3$\cite{Machado2003}, $^4$\cite{Cidale2012}, $^5$\cite{Swings1981}, $^6$\cite{Millour2011}, $^7$\cite{Muratore2012}, $^8$\cite{Aret2016}, $^9$\cite{Oksala2013}, $^{10}$\cite{Muratore2015}, $^{11}$\cite{Mehner2016}, $^{12}$\cite{DomicianodeSouza2011}]}

\section{Discussion-Conclusions}

By inspecting the structures finally derived we notice two cases with respect to the innermost ring. In HD 327083, MWC 137, and CPD-57 2874 the CO ring is the innermost ring, i.e. there is no indication of atomic gas (at least under the proper conditions to produce detectable emission lines) further inside that ring. On the contrary, there are regions of atomic gas further outside. On the other hand, in HD 62623, CPD-52 9243, and Hen 3-298 there is atomic gas closer than the CO ring. It can be also found further out or even coexisting with molecular gas. These structures seem more consistent with a picture of a radial and outwards decrease of density and temperature.

Binarity could be responsible for "clearing" the atomic gas and resulting in the formation of the CO ring as the first region closer to the system. This is true in the case of HD 327083 where CO has been found to be circumbinary, but HD 62623 contradicts this paradigm. It is also a binary system in which there is presence of atomic gas inside the CO ring. Unless we manage to gather more data on the binarity among the B[e]SGs, its role will remain elusive. 

Even though we cannot derive a common picture for the exact structure of the circumstellar environment around these B[e]SGs, as each one of them has its own character, we do see that they consist of a sequence of multiple rings. This should be the result of a common formation mechanism, such as mass loss triggered by non-radial pulsations and/or other  instabilities, or even due to the presence of objects that can clear their paths and stabilize these ring structures similar to shepherd moons \citep{Kraus2016}. Even after 40 years of studies, there are still open and puzzling questions.

\acknowledgements GM and MK acknowledge financial support from GA \v{C}R (14-21373S); AA from Estonia grant IUT40-1.
The Astronomical Institute Ond\v{r}ejov is supported by the project RVO:67985815. Observations obtained from 2014 to 2016 with the MPG 2.2m telescope were supported by the Ministry of Education, Youth and Sports project - LG14013 (Tycho Brahe: Supporting Ground-based Astronomical Observations). Some observations also made with the MPG 2.2 m telescope under the agreement between ESO and Observat\'orio Nacional/MCTI and Max Planck Institut f\"ur Astronomie and Observat\'orio Nacional/MCTI (Brazil). 

\bibliography{/home/grigoris/Resources/Papers-Library/00_bibs_00/supergiants}

\end{document}